\documentclass[runningheads]{svmult}
\usepackage{makeidx}   
\usepackage{graphicx}  
\usepackage{subeqnar}  
\usepackage{multicol}  
\usepackage{cropmark} 
\usepackage{physprbb}  
\begin{document}
\title*{Optimal Supernova Search Strategies}
%
%
%
%
\titlerunning{Optimal Supernova Search Strategies}
%
\author{Dragan Huterer\inst{1}
\and Michael S. Turner\inst{1,2,3}}
\authorrunning{D. Huterer and M.S. Turner}
%
%
\institute{Department of Physics, University of Chicago, Chicago IL 60637, USA
\and Department of Astronomy and Astrophysics, University of Chicago, Chicago IL 60637, USA
\and Fermi National Accelerator Laboratory, Batavia IL 60510-0500, USA}

\maketitle              

\begin{abstract}
Recent use of Type Ia supernovae to measure acceleration of the
universe has motivated questions regarding their optimal use to
constrain cosmological parameters $\Omega_M$, $\Omega_{\Lambda}$ and
$w_Q$.  In this work we address the question: what is the optimal
distribution of supernovae in redshift in order to best constrain the
cosmological parameters?  The solution to this problem is not only of
theoretical interest, but can be useful in planning supernova
searches.  Using the Fisher matrix formalism we show that the error
ellipsoid corresponding to $N$ parameters (for $N=1, 2, 3$) has the
smallest volume if the supernovae are located at $N$ discrete
redshifts, with equal number of supernovae at each redshift and with
one redshift always being the maximum one probed.  Including
marginalization over the ``nuisance parameter'' $\mathcal{M}$ changes
this result only trivially.

\end{abstract}
\section{Introduction}
Recent use of Type Ia supernovae (SNe Ia) as standard candles
\cite{perlmutter-1997,perlmutter-1999,schmidt,riess} provided an
opportunity to obtain the distance-redshift relation -- and measure
the cosmological parameters -- without recourse to the ``distance
ladder''. SNe Ia provide strong evidence for the acceleration of the
universe, thus implying that a component called ``dark energy'' with
strongly negative pressure ($p< -1/3\,\rho$) dominates the
energy-density of the universe.  The most likely candidate for the
dark energy is the vacuum energy, with $p=-\rho$. SNe Ia provide the
luminosity distance -- redshift relation and thus effectively
determine the parameters $\Omega_M$ and $\Omega_{\Lambda}$, energy
densities in matter and vacuum energy scaled to the critical energy
density. The  best-fit value using the current dataset of $\sim 50$ SNe
Ia and assuming a flat universe is $\Omega_M=1-\Omega_{\Lambda}\simeq
0.3$, with $1$-$\sigma$ error of order $0.1$.

Given the importance of SNe Ia as standard candles --- namely,
determining the contents of the universe and probing the dark energy
--- it is important to consider what can be done to improve upon
the constraints on cosmological parameters. Two obvious improvements
would be increasing the size of the supernova sample and better
control over the systematic errors.  Going to redshifts beyond one is
also very high on supernova cosmologists' wish list because that way
one can differentiate between the effects of dust and dark energy.  It
is believed that, barring unusual scenarios, dust would make faraway
objects progressively fainter, while the dark energy becomes
inoperative at $z> 1$ because the universe is essentially
matter-dominated at those redshifts. Finally, an important requirement
would be to have supernovae located throughout the probed redshift
range, in order to make sure that the magnitude-redshift relation
traced out corresponds to the dark energy, and not evolution or dust.
The planned supernova space telescope, SNAP~\cite{snap-page}, would
satisfy all of the above-mentioned requirements.
 
This work considers supernova search strategies for the most accurate
determination of cosmological parameters $\Omega_M$ and
$\Omega_{\Lambda}$ (and possibly the equation of state ratio of the
dark energy, $w_Q$, where 'Q' stands for 'quintessence'
\cite{quint}). To this end, we ask: given the cosmological parameters
we want to determine, what is the optimal distribution of supernovae
in redshift in order to best constrain those parameters? At first
glance this problem may appear of purely academic interest since we
are not free to put supernovae where we please.  However, supernova
observers have considerable freedom in choosing redshift ranges for
their searches, by using filters sensitive to wavelengths
corresponding to spectra at observed  redshifts.  Moreover,
the increased difficulty in observing high-redshift supernovae means
that, even with great improvement in supernova detection and follow-up
techniques, it can be as time-consuming to observe one supernova at,
say, $z=1.4$ as many $z<1$ supernovae. Hence, an observer will have to
decide how much telescope time is to be allocated to specific redshift
ranges to best constrain the cosmological parameters.

In this work we make three assumptions:

$\bullet$ Magnitude uncertainty, $\sigma_m$, is the same for each supernova
irrespective of redshift (this is a pretty good approximation, at
least for current data sets).

$\bullet$  Total number of observed supernovae is fixed (rather than
the total observing time, for example).

$\bullet$  There is an unlimited number of supernovae at each
redshift.

None of these assumptions is required to use the formalism we present. 
Moreover, the results we present should qualitatively not be very
different from those obtained when the constraints above are relaxed.
We make the assumptions above to illustrate our approach and simplify the
analysis.

\section{The Most Accurate Parameter Determination}

We tackle the following problem: given $n$ supernovae and their
corresponding uncertainties, what distribution of these supernovae in
redshift would enable the most accurate determination of cosmological
parameters? In case of more than one parameter, we need to define
what we mean by most accurate determination of all parameters
simultaneously.  Since the uncertainty in measuring $N$
parameters simultaneously is described by an $N$-dimensional ellipsoid
(at least under assumption that the total likelihood function is
gaussian), we make a fairly obvious and, as it turns out,
mathematically tractable requirement that the ellipsoid have minimal
volume.  This requirement corresponds to the best {\it local}
determination of the parameters.

We now show that volume of the ellipsoid is given by 

\begin{equation}  
V \propto \det(F)^{-1/2},
\label{vol_ellipsoid}
\end{equation}
where the sign of proportionality expresses our ignorance of
a numerical factor, and $F$ is the Fisher matrix
\cite{Fisher-Jungman,Fisher-Tegmark}

\begin{equation}
F_{ij} = -\left< \partial^2 \ln L \over \partial p_i \partial p_j
 \right>_{\bf  y},
\end{equation}
where $L$ is the likelihood of observing data set ${\bf y}$ given the
parameters $p_1 \ldots p_N$. Although this relation might be
familiar/obvious to mathematically inclined cosmologists, we present its
derivation for completeness.

To prove equation (\ref{vol_ellipsoid}), consider a general
uncertainty ellipsoid in $n$-dimensional parameter space. 
The equation of this ellipsoid is
\begin{equation}
X^T F X =1,
\end{equation}
where $X=(x_1 x_2 \ldots x_N)$ is the vector of coordinates and
$F$ the Fisher matrix. Let us now rotate the ellipsoid so that
all of its axes are parallel to coordinate axes.  Equivalently we
can rotate the coordinates to achieve the same effect, by writing
$X_{\rm rot}=UX$, where $U$ is the orthogonal matrix
corresponding to this rotation. The equation of the ellipsoid in
the new coordinate system is
\begin{equation}
X_{\rm rot}^T \,F\, X_{\rm  rot} =1,
\end{equation}
or  equivalently, in the original coordinate system 
\begin{equation}
X^T \,F_{\rm rot}\, X=1,
\end{equation}
where $F_{\rm rot}=U^T F U$ is the Fisher matrix for the rotated
ellipsoid, and has the form $F_{\rm par }=diag(1/\sigma_1^2, \ldots, 1/\sigma_N^2)$
with $\sigma_i$ the i-th axis of the ellipsoid. The volume of the
(rotated) ellipsoid is obviously

\begin{equation}
V\propto \prod_{i=1}^N \sigma_i = \det(F_{\rm rot})^{-1/2}.  
\end{equation}

Then, since
$\det(F)=\det(F_{\rm rot})$ and rotations preserve volumes, we have 
\begin{equation}
V \propto \det(F_{\rm rot})^{-1/2}= \det(F)^{-1/2},
\end{equation}
and this completes  the proof.

\subsection{Fisher Matrix for Supernovae}

To minimize the volume of the ellipsoid we therefore need to
maximize $\det(F)$.  Fisher matrix for the case of supernova
measurements was first worked out by Tegmark et al.~\cite{CMB+SN},
and we briefly recapitulate their results, with slightly
different notation and one addition. The measurements
are given as 
\begin{equation}
m_n=5\log\;[H_0 d_L(z_n, \Omega_M, \Omega_\Lambda)]+{\mathcal{M}}+\epsilon_n,
\end{equation}
where $m_n$ is apparent magnitude of the $n$th supernova in the
sample, $d_L$ is its luminosity distance,
${\mathcal{M}}\equiv  M-5\log(H_0)+25$~\cite{perlmutter-1999} (with $M$
absolute magnitude of a supernova), and $\epsilon_n$ is the error
in that measurement (assumed to be drawn from a gaussian
distribution with zero mean and standard deviation
$\sigma_m$). Note that ${\mathcal{M}}$ contains all dependence on
$H_0$, since $d_L\propto 1/H_0$.

The Fisher matrix is given by
\cite{CMB+SN}

\begin{equation}
F_{ij}=\frac{1}{\sigma_m^2} \sum_{n=1}^{N} w_i(z_n)
w_j(z_n)
\end{equation}
where $w$'s are weight functions given by
\begin{equation}
w_i(z) \equiv 
{5\over\ln 10}
\left\{{\kappa S'[\kappa I(z)]\over S[\kappa I(z)]}
\left[{\partial I\over\partial p_i} - { I(z)\over  2\kappa^2}\right]
+ {1\over  2\kappa^2}
\right\},  \label{w_defined}
\end{equation}
if the parameter $p_i$ is $\Omega_M$ or $\Omega_{\Lambda}$ 
(or $\Omega_Q$), or else
\begin{equation}
w_i(z) \equiv 
{5\over\ln 10}
\left[{\kappa S'[\kappa I(z)]\over S[\kappa I(z)]} \,
{\partial I\over\partial p_i}\right]
\end{equation}
if $p_i$ is $w_Q$. Also

\begin{equation}
H_0\,d_L = (1+z)\, \frac{S(\kappa I)}{\kappa},
\end{equation}

\begin{equation}
S(x)=\left \{ \begin{array}{cl} \sinh(x), & \;\mbox{if}\;\; \Omega_{\rm TOT} > 1;\\ 
                                {x}      , & \;\mbox{if}\;\; \Omega_{\rm TOT} = 1;\\  
 		                \sin(x) , & \;\mbox{if}\;\; \Omega_{\rm TOT} <  1. 
              \end{array} \right. 
\end{equation}
\vspace{-0.2cm}
\begin{eqnarray}
I(z, \Omega_M, \Omega_{ \Lambda}) &=& \int_0^z A(z')^{-1/2} dz' \\[0.1cm] 
A(z) &=& \Omega_M (1+z)^3 + \Omega_{\Lambda} + \kappa^2 (1+z)^2
\label{A_def} \\ [0.1cm] 
\kappa^2  &=& 1- \Omega_M - \Omega_{\Lambda}.
\end{eqnarray}
Here $\Omega_M$, $\Omega_{\Lambda}$ and $\kappa^2$ are the energy
densities in matter, cosmological constant and curvature
respectively divided by the critical density, and $\Omega_{\rm
TOT}\equiv \Omega_M + \Omega_{\Lambda}$.  Later on we also
consider a more general equation of state for the exotic energy,
and replace the second term in equation (\ref{A_def}) by $\Omega_Q
(1+z)^{3(1+w_Q)}$.

In addition to $\Omega_M$, $\Omega_Q$ and $w_Q$, the
magnitude-redshift relation also includes the ``nuisance parameter''
$\mathcal{M}$, which is a combination of the Hubble parameter and
absolute magnitude of supernovae, and which has to be marginalized
over in order to obtain constraints on parameters of
interest. Ignoring $\mathcal{M}$ in the Fisher matrix formalism (that
is, assuming that $\mathcal{M} $ is known) leads to a serious
underestimate of the uncertainties in other parameters (of course,
fairly accurate knowledge of $H_0$ could be used to obtain $M$ from a
sample of nearby supernovae, and thus determine $\mathcal{M}$).  We
continue to ignore $\mathcal{M}$ for simplicity and mathematical
clarity, and in section~\ref{mathcal-sec} we show that including
marginalization over $\mathcal{M}$ changes our results rather
trivially.

The Fisher matrix can further be written as
\begin{eqnarray}
F_{ij} &=& {N\over\sigma_m^2}\int_0^{z_{\rm max}} g(z)  w_i(z)
w_j(z) dz, \label{F_ij}
\end{eqnarray}
where
\begin{equation}
g(z) = {1\over N}\sum_{n=1}^N \delta(z-z_n) 
\label{define_g}
\end{equation}
is the (normalized) distribution of redshifts of the data and $z_{\rm
max}$ is the highest redshift probed in the survey.  {\it Our goal
is to find $g(z)$ such that $\det(F)$ is maximal}.  $g(z)$ is
essentially a histogram of supernovae normalized to have unit
area. Note that the maximization of $\det(F)$ will not depend on $N$
and $\sigma_m$, so we drop them from further analysis. To consider a
non-constant error $\sigma_m(z)$, one simply absorbs $\sigma_m(z)$
into the definition of weight functions $w(z)$.

\section{Results}\label{sec-results}

\subsection{Determination of One Parameter}

We first consider the trivial -- but instructive -- case of one
parameter. Then we need to maximize $\int_0^{z_{\rm max}} g(z)\,
w_1^2(z)\,dz$, subject to $ \int_0^{z_{\rm max}} g(z) \, dz=1$ and
$g(z)\geq 0$.  It is quite obvious that the solution is a single
delta function at the redshift where $w_1(z)$ has a maximum. For
any given parameter $w_1(z)$ will have a maximum at $z_{\rm max}$ and that
is where we want all our supernovae to be. This result is hardly
surprising: we have a one-parameter family of curves $m(z)$, and
the best way to distinguish between them is to have all
measurements at the redshift where the curves differ the most, at
$z_{\rm max}$.

\begin{figure}[h]
\begin{center}
\includegraphics[width=3.5in, height=2.5in]{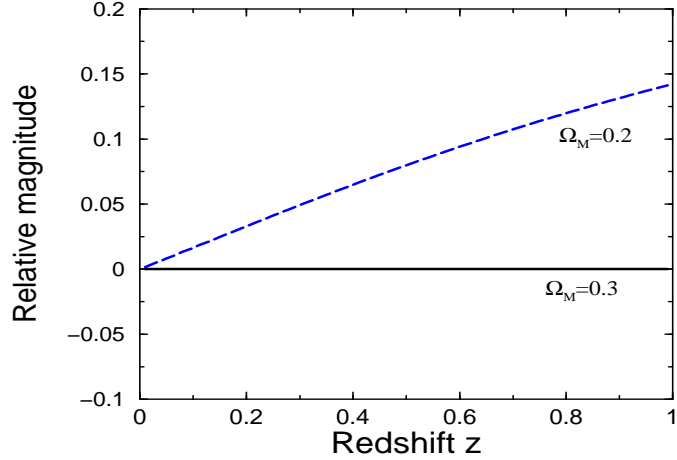}
\end{center}
\caption[]{Dependence of the magnitude-redshift curve on a single
cosmological parameter. We assume a flat universe and vary
$\Omega_M=1-\Omega_{\Lambda}$.  Magnitudes for the $\Omega_M=0.3$ flat
model were subtracted from both curves for easier visualization }
\label{magdiff.fig}
\end{figure}

As an example, Fig.~\ref{magdiff.fig} shows magnitude-redshift
curves for the fiducial $\Omega_M=0.3$ model with the assumption
$\Omega_{\Lambda}=1-\Omega_M$ (flat universe). As $\Omega_M$ is varied, the
biggest difference in $m(z)$ is at the highest redshift
probed. In order to best constrain $\Omega_M$, therefore, all
supernovae should ideally be located at $z=1.0$, our assumed maximum redshift.

\subsection{Determination of Two Parameters}

\begin{figure}[h]
\begin{center}
\begin{minipage}[t]{6.5in}
\includegraphics[width=2.4in, height=2in]{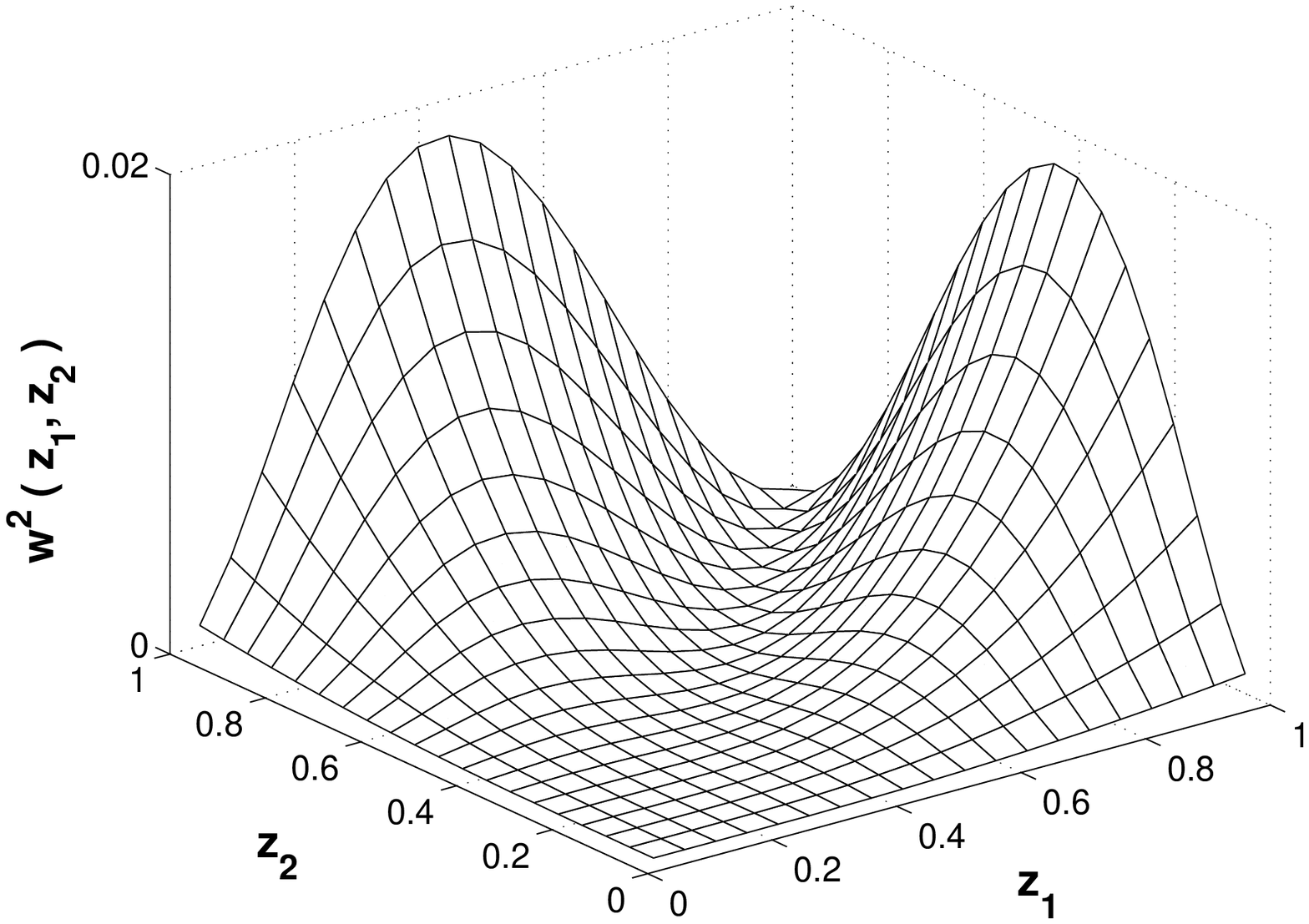}
\begin{minipage}[t]{6.5in}
\includegraphics[width=2.4in, height=2in]{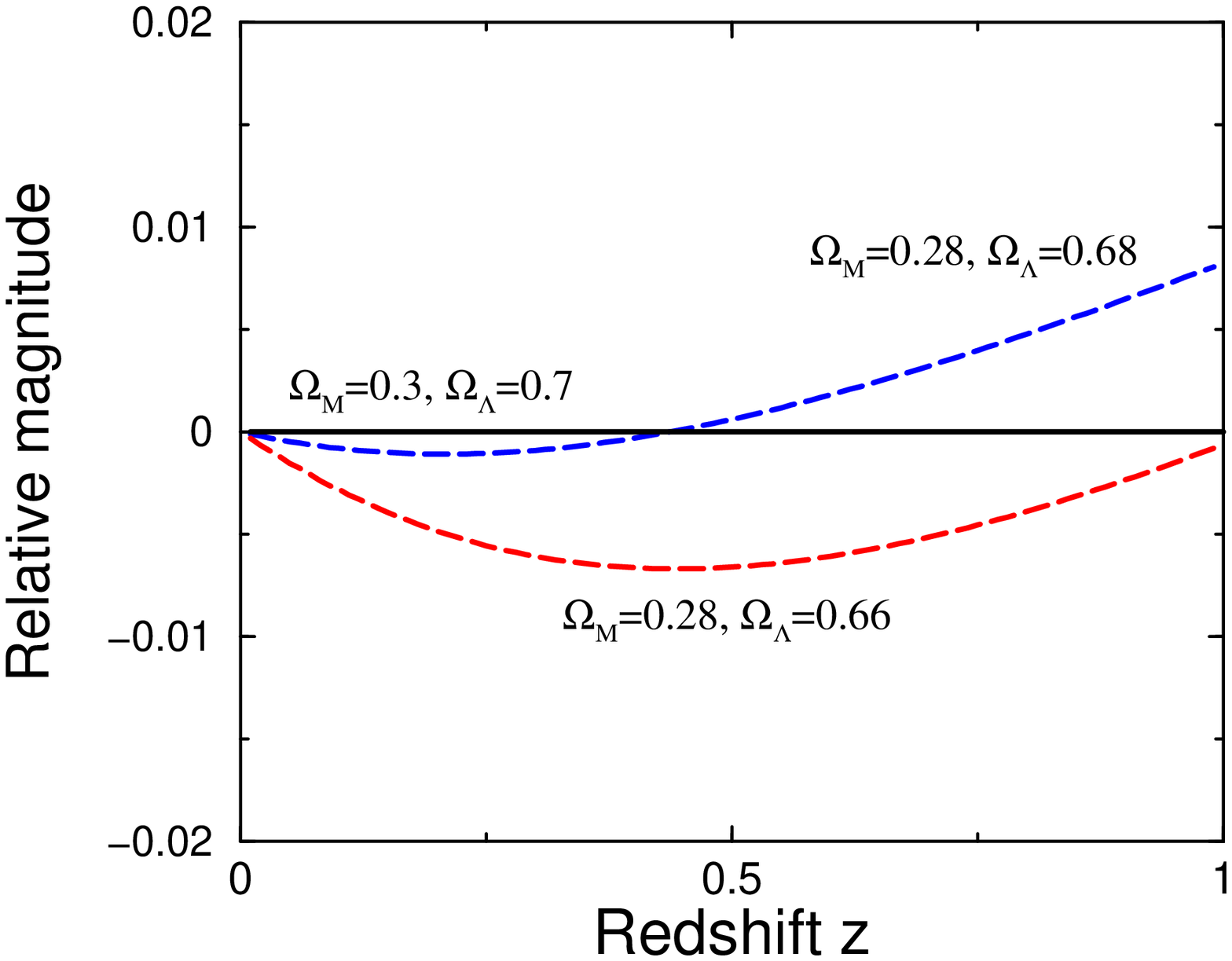}
\end{minipage}
\end{minipage}
\end{center}
\caption[]{{\it Left panel:} Function $w^2 {(z_1, z_2)}$ for the case
when $\Omega_M=0.3$ and $\Omega_{\Lambda}=0.7$. {\it Right panel:}
Variation of the magnitude-redshift curve when two cosmological
parameters, $\Omega_M$ and $\Omega_{\Lambda}$, are varied independently }
\label{weight.fig}
\end{figure}

A more interesting -- and relevant -- problem is minimizing the
area of the ellipse which describes the uncertainties for two
parameters. The expression to maximize is then

\begin{eqnarray}
 && \int_{z=0}^{z_{\rm max}} g(z) \,w_1^2(z) dz 
\int_{z=0}^{z_{\rm max}} g(z) \, w_2^2(z)  dz\;  - 
\left (\int_{z=0}^{z_{\rm max}} g(z) \,w_1(z)\,  w_2(z) dz\right
)^2  \nonumber \\[0.2cm] 
&=& {1\over 2} \int_{z_1=0}^{z_{\rm max}}\int_{z_2=0}^{z_{\rm
max}} g(z_1)\, g(z_2) 
\,w^2(z_1, z_2) \,dz_1\, dz_2, \label{area_analytic}
\end{eqnarray}
where $w(z_1, z_2)\equiv w_1(z_1)w_2(z_2) -
w_1(z_2)w_2(z_1)$ is a known function of redshifts and
cosmological parameters (see Fig.~\ref{weight.fig}) and $g(z)$ is subject to the same
constraints as in case of one parameter.

Despite a relatively harmless appearance of expression
(\ref{area_analytic}),
we found it impossible to maximize it analytically. Fortunately, for all
practical purposes numerical solution will be
sufficient. Returning to the discretized version of equation
(\ref{F_ij}), we divide the interval $(0, z_{\rm max})$ into $B$ bins with
$g_i N$ supernovae in bin $i$. Then we need to maximize
\begin{equation}
\sum_{i, j=1}^{B} g_i\,g_j\, w^2(z_i, z_j) \label{g_ij}
 \end{equation} 
subject to
\begin{equation}
\sum_{i=1}^{B} g_i=1 \,\,\,\,\, \mbox{and} \,\,\,\,\ g_i \geq 0. 
\label{quadprog.constr}
\end{equation}

Equations (\ref{g_ij}) and (\ref{quadprog.constr}) define a quadratic
programming problem --- extremization of a quadratic function subject
to linear constraints. Since $w^2(z_1, z_2)$ is neither concave nor
convex (see Fig.~\ref{weight.fig}), we have to resort to brute force
maximization, and consider all possible values of $g_i$. 
The result of this maximization is that the
optimal distribution is two delta functions of equal magnitude:
\begin{equation}
g(z)= 0.50\; \delta(z-0.43) + 0.50 \; \delta(z-1.00) ,
\end{equation}
where all constants are accurate to $0.01$. Thus, half of the
supernovae should be at the highest available redshift, while the 
other half at about 2/5 of the maximum redshift. 

This result is not very sensitive to the maximum redshift probed, or
fiducial parameter values. If we increase the maximum available redshift to
$z_{\rm max}=1.5$, we find two delta functions of equal magnitude at
$z=0.57$ and $z=1.50$. If we change the fiducial values of
parameters to $\Omega_M=0.3$ and $\Omega_{\Lambda}=0$ (open
universe), we find delta functions of equal magnitude at $z=0.47$ 
and $z=1.00$.

Finally, let us consider a different choice for the two parameters --- for
example, $\Omega_M$ and $w_Q$ (equation of state of
the dark energy), with fiducial values  $\Omega_M=0.3$ and $w_Q=-1$
and with the assumption of flat universe
($\Omega_Q=1-\Omega_M$).  Then, assuming $z_{\rm max}=1.0$, we get  
\begin{equation}
g(z)= 0.50 \;\delta(z-0.36) + 0.50\; \delta(z-1.00),
\end{equation}
and again optimal distribution of supernovae is similar to the case of
$\Omega_M$ and $\Omega_{\Lambda}$ as parameters.

\subsection{Determination of Three Parameters}

We further  consider parameter determination with three parameters
$\Omega_M$, $\Omega_Q$ and $w$. Elements of the 3x3 Fisher
matrix are calculated according to expression (\ref{F_ij}), and we
again maximize $\det(F)$ as  described above. The result is 
\begin{eqnarray}
g(z)&=& 0.33 \;\delta(z-0.21) + 0.34\; \delta(z-0.64) + \nonumber \\
 && 0.33\;\delta(z-1.00),
\end{eqnarray}
with all constants accurate to 0.01.  Hence we have three
delta functions of equal magnitude, with one of them at the
highest available redshift.

It appears impossible to prove that $N$ parameters are best measured
if the data form $N$ delta functions in redshift. However, it is easy
to prove that, {\em if} the data do form $N$ delta functions, then
those delta functions should be of equal magnitude and their locations
should be at coordinates where the ``total'' weight function
(e.g. $w^2(z_1, z_2)$ in case of two parameters) has a global
maximum.  In practice, the relevant number of cosmological parameters
to be determined from the SNe Ia data is between one and three, so
considering more than three parameters is less relevant for practical
purposes.

\subsection{Including Marginalization over $\mathcal{M}$}\label{mathcal-sec}

So far we have been ignoring the parameter $\mathcal{M}$, assuming
that it is known (equivalently, that the value of $H_0$ and the absolute
magnitude of supernovae are precisely known). This, of course, is not
the case, and $\mathcal{M}$ is marginalized over to obtain
probabilities on cosmological parameters. Fortunately, when
$\mathcal{M}$ is properly included, our results change in a
predictable and rather trivial way, as we now show.

\begin{figure}[ht]
\begin{center}
\includegraphics[width=3.5in, height=2.5in]{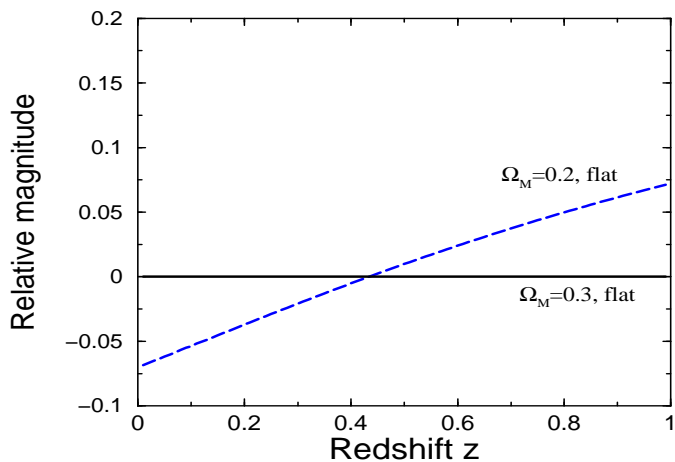}
\end{center}
\caption[]{ Magnitude-redshift curve when a single cosmological
parameter, $\Omega_M$, is varied. Marginalization over $\mathcal{M}$
is now included, which means that the curves are allowed to slide
vertically as well }
\label{magdiff.offset.fig}
\end{figure}

Including $\mathcal{M}$ as an undetermined parameter, we now have
an $N+1$-dimensional ellipsoid ($N$ cosmological parameters plus
$\mathcal{M}$), and we want to minimize the volume of its
projection onto the $N$-dimensional space of cosmological
parameters. The equation of the $N$-dimensional projection
is

\begin{equation}
X^T F_{\rm proj} X=1
\end{equation}
and $F_{\rm proj}$ is obtained as follows: 1) Invert the original
$F$ to obtain the covariance matrix $F^{-1}$\hspace{0.1cm} 2)
pick the desired $N$x$N$ subset of $F^{-1}$ and call it $F_{\rm
proj}^{-1}$, and \hspace{0.08cm} 3) invert it to get $F_{\rm
proj}$.

Minimizing the volume of the projected ellipsoid we obtain the result
that the optimal supernova distribution is obtained with $N$ delta
functions in redshift obtained when ignoring $\mathcal{M}$, plus the
delta function at $z=0$. All $N+1$ delta functions have the
same magnitude. The intuitive explanation for this result is
illustrated in Fig.~\ref{magdiff.offset.fig}, which shows the
magnitude-redshift curves when $\Omega_M$ and $\mathcal{M}$ are varied
(flat universe is assumed). This figure is the same as
Fig.~\ref{magdiff.fig}, except the curves are now allowed to slide
vertically as well, corresponding to the variation in $\mathcal{M}$. 
There are two locations of largest
departure when the two parameters are varied, namely $z=0$ and
$z=1$. It makes sense then that those are precisely the locations
where the supernovae should be, and our analysis says that we ideally
need equal number of supernovae at each location.

\subsection{Optimal vs Uniform Distribution}

\begin{figure}[ht]
\begin{center}
\begin{minipage}[t]{6.5in}
\includegraphics[width=2.4in, height=2in]{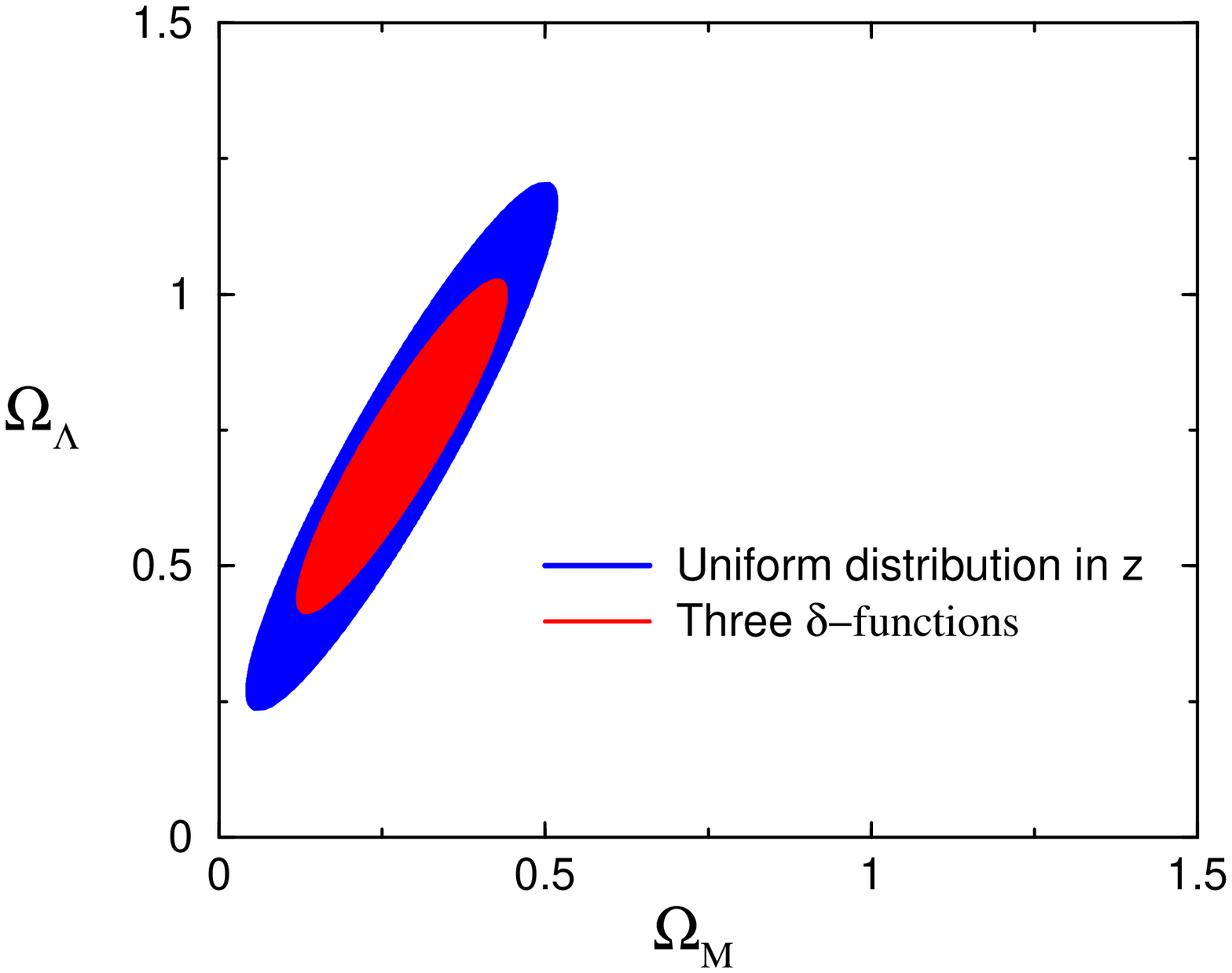}
\begin{minipage}[t]{6.5in}
\includegraphics[width=2.4in, height=2in]{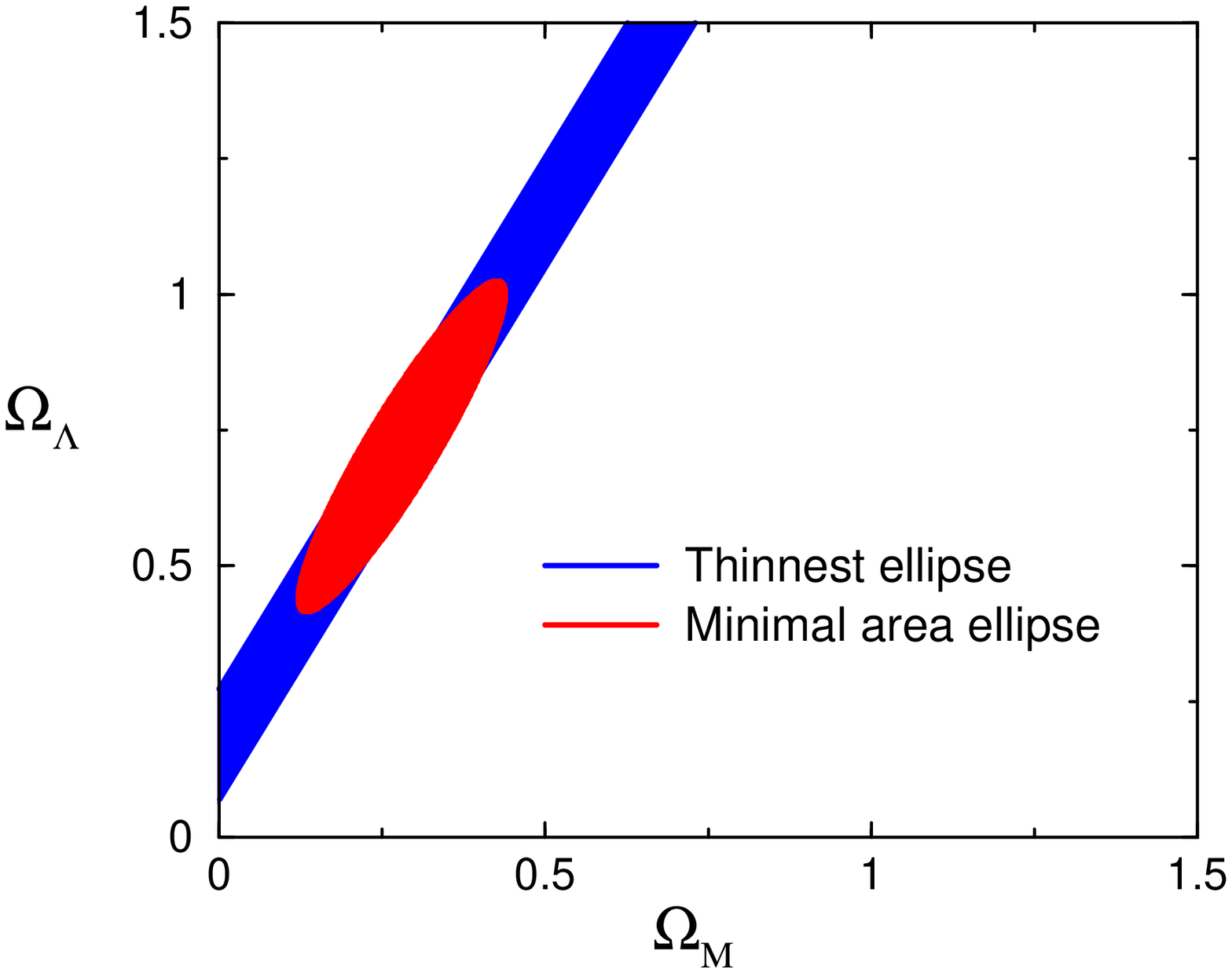}
\end{minipage}
\end{minipage}
\end{center}
\caption[]{ {\it Left panel:} Uniform (dark-blue) vs optimal
(light-red) distribution in redshift.  Shown are constraints on
$\Omega_M$ and $\Omega_{\Lambda}$ when $\mathcal{M}$ was marginalized
over.  To get the absolute sizes of the ellipses, $N=100$ SNe were
assumed with individual uncertainties of $\sigma_m=0.15$ mag. {\it
Right panel:} Thinnest possible ellipse for given $N$ and $\sigma_m$
(dark-blue) is infinitely long in one direction. However, the
smallest-area ellipse (light-red) is almost as thin, and for all
purposes can serve as the thinnest ellipse }
\label{compare.ellipses.fig}
\end{figure}

Are the advantages of the optimal distribution significant enough that
one should consider them seriously? In our opinion the answer is
affirmative, as we illustrate in the left panel of
Fig.~\ref{compare.ellipses.fig}.  This figure shows that the area of
the $\Omega_M$-$\Omega_{\Lambda}$ uncertainty ellipsoid is more than
two times smaller if the SNe have the optimal distribution in redshift
as opposed to the case of uniform distribution.

\subsection{Requiring the {\it Thinnest} Ellipse}

If we are using  SNe Ia alone to determine the cosmological
parameters, then we clearly want to minimize the area of the error
ellipse (we consider the case of $\Omega_M$ and $\Omega_{\Lambda}$ as
parameters in this section). However, supernova
measurements will also be combined with other methods to determine the
parameters.  A prime example of a symbiosis between two or more
methods is combining CMB measurements with those of supernovae
\cite{Zaldarriaga,CMB+SN}. These methods together can improve the
determination of $\Omega_M$ and $\Omega_{\Lambda}$ up to a factor of
10 as compared to either method alone due to breaking of the
degeneracy between the two parameters. As can be seen in Fig.~2 of
ref.~\cite{CMB+SN}, in combining the CMB with SNe Ia data one
might hope for the {\it thinnest} ellipse possible coming from
supernova measurements. Here by ``thin'' we mean that the combination
$\Omega_M - \Omega_{\Lambda}$ is accurately determined.

Finding the thinnest ellipse is the problem that we can solve using
our formalism, since the length of each axis of the ellipse is
proportional to the inverse square root of an eigenvalue of the
corresponding Fisher matrix. All we need to do then is maximize the
larger eigenvalue of $F$ with respect to the distribution of the
supernovae $g(z)$.

The result is perhaps not at all surprising: to get the thinnest
ellipse, all supernova measurements should be at the same
(maximum) redshift, which at the same time implies an infinitely
long ellipse. More generally, we find that changing the redshift
distribution of supernovae doesn't change the thickness of the
error ellipse greatly, but does change its length. By attempting
to obtain a thinner ellipse, we end up only making it
longer. Fortunately, the {\em smallest area} ellipse we found can
in practice serve as the thinnest ellipse, as shown in
the right panel of Fig.~\ref{compare.ellipses.fig}.

\subsection{Reconstruction of the Potential of Quintessence}

It has been shown recently that a sufficiently good supernova
sample could be used to reconstruct the potential of quintessence
out to $z\simeq 1$ \cite{reconstr,chiba}. More generally,
equation of state ratio of the missing energy, $w_Q$, can also be
reconstructed
\cite{reconstr}.

In the spirit of our analyses above, we ask: what distribution of 
supernovae in redshift gives the smallest 95\% confidence region 
for the reconstructed potential $V(\phi)$? To answer this
question, we perform a Monte-Carlo simulation by using different
distributions of supernovae and computing the average area of the 
confidence region corresponding to each of them. 

Uniform distribution of supernovae gives the best result among the
several distributions we put to test. This is not surprising, because
reconstruction of the potential consists in taking first and second
derivatives of the distance-redshift curve, and the most accurate
derivatives are obtained if the points are distributed uniformly.  For
comparison, gaussian distribution of supernovae with $\overline{z}
=0.7, \sigma_z=0.4$ gives the area that is $10-20$\% larger.

\section{Discussion} \label{sec-discussion}

We considered supernova search strategies that produce the tightest
constraint on cosmological parameters by minimizing the volume of the
error ellipsoid corresponding to those parameters.  We first proved
that, assuming that the total likelihood function is gaussian, the
volume of an $N$-dimensional error ellipsoid is proportional to the
inverse square root of the determinant of the corresponding Fisher
matrix. Then we showed that, if the supernova measurements are used to
determine $N$ parameters with $N=1, 2$ or $ 3$, this volume is
minimized if the distribution of supernovae in redshift is given by
$N$ delta functions of equal magnitude.  In particular, $\Omega_M$ and
$\Omega_{\Lambda}$ have the smallest error ellipse if half of the
measurements are at $z_{\rm max}$, while the other half are at roughly
$2/5\, z_{\rm max}$. If $\mathcal{M}$ is marginalized over, we need an
additional supernova sample at $z=0$ (a total of $N+1$ delta functions
in redshift).

Our approach is quite flexible and can be applied in practice. 
For example, given the redshift-dependence of the
measurement uncertainties for a given experiment, $\sigma_m(z)$,
as well as the time it takes to obtain a measurement at a given
redshift (determined by the telescope specifications), one can perform
analysis similar to that in section \ref{sec-results} to infer
the optimal search strategy. 

It is important to keep in mind that the best determination of the
parameters around their fiducial values is only one of the possible
objectives. As mentioned in the introduction, tracing out the
magnitude-redshift curve throughout the redshift range is important
to test for systematics such as evolution or dust.  A sample of nearby
supernovae would be useful to further check for systematics (at least
one low-z SN search program is already under way). Ultimately, the
chosen supernova search strategy should combine all of these
considerations.

\paragraph{Acknowledgments.} We would like to thank members of the Supernova 
Cosmology Project and Daniel Holz for useful discussions.  This work
was supported by the DoE (at Chicago and Fermilab) and by the NASA (at
Fermilab by grant NAG 5-7092).

%

\end{document}